\documentclass[conference]{IEEEtran}
\IEEEoverridecommandlockouts
\usepackage{cite}
\usepackage{amsmath,amssymb,amsfonts}
\usepackage{url}
\usepackage{hyperref}
\usepackage{breakurl}
\usepackage{algorithmic}
\usepackage{graphicx}
\usepackage{textcomp}
\usepackage{xcolor}
\usepackage{comment}
\usepackage{rotating}

\def\BibTeX{{\rm B\kern-.05em{\sc i\kern-.025em b}\kern-.08em
    T\kern-.1667em\lower.7ex\hbox{E}\kern-.125emX}}
    
\begin{document}

\title{ Real-Time Monitoring and Transparency in Pizza Production Using IoT and Blockchain \\ }

\author{
    \IEEEauthorblockN{
        Azmat Ullah$^{1,2}$,  
        Maria Ilaria Lunesu$^{2}$,  
        Lodovica Marchesi$^{2}$,
        Roberto Tonelli$^{2}$
    }
    \IEEEauthorblockA{
        $^{1}$University of Camerino and 
        $^{2}$University of Cagliari\\
        azmat.ullah@unicam.it, mariai.lunesu@unica.it, lodovica.marchesi@unica.it, roberto.tonelli@unica.it
    }
}

\maketitle

\begin{abstract}
This paper presents a blockchain-based Internet of Things (IoT) system for monitoring pizza production in restaurants. IoT devices track temperature and humidity in real-time, while blockchain ensures secure and tamper-proof data. A Raspberry Pi processes sensor data, captures images, triggers alerts, and interacts with smart contracts. The system detects abnormal conditions, enabling quick responses. Blockchain adds transparency and traceability, supporting compliance and audits. Experiments show improved ingredient management, reduced waste, and increased kitchen efficiency.
\end{abstract}


\section{Introduction}

Maintaining ingredient quality and safety is critical in pizza production, where fresh components like dough and cheese are highly sensitive to temperature changes. As the global pizza market grows, efficient environmental monitoring becomes essential.

This work proposes a smart system integrating IoT sensors and blockchain technology to monitor temperature and humidity in real time. A Raspberry Pi processes the data, triggering alerts for threshold breaches and ensuring secure, tamper-proof logging via smart contracts.

The system enables data-driven decisions, reduces waste, and ensures compliance with HACCP food safety standards. Designed for real-world pizzerias, it offers a scalable, energy-efficient solution that enhances operational reliability and can be extended to the broader food industry.

\section{Related Work}
\label{related}

The combination of IoT and blockchain has been widely explored to enhance food safety, environmental monitoring, and supply chain transparency. IoT enables real-time tracking of temperature and humidity, while blockchain ensures secure, immutable data storage, improving traceability and operational reliability~\cite{K1-3265692}. Existing studies have demonstrated their effectiveness in managing perishable goods across agriculture, logistics, and retail~\cite{Addou2023}, supporting sustainability and reducing food waste~\cite{agriculture13061173}. Platforms like Hyperledger and Ethereum provide scalable solutions, while architectures such as SHEEPDOG ensure data privacy and integrity~\cite{9156150}. Despite proven benefits, adoption in small restaurants remains limited due to cost and complexity~\cite{blockchains2030016}. Most implementations focus on large-scale systems, leaving a gap in niche areas like pizza production.
This work addresses that gap by proposing an affordable, energy-efficient solution tailored to pizzerias, integrating real-time monitoring and blockchain traceability to improve food safety and ingredient management.

\section{System Design and Implementation}



\begin{figure}[h] 
  \centering  
  \includegraphics[width=\columnwidth]{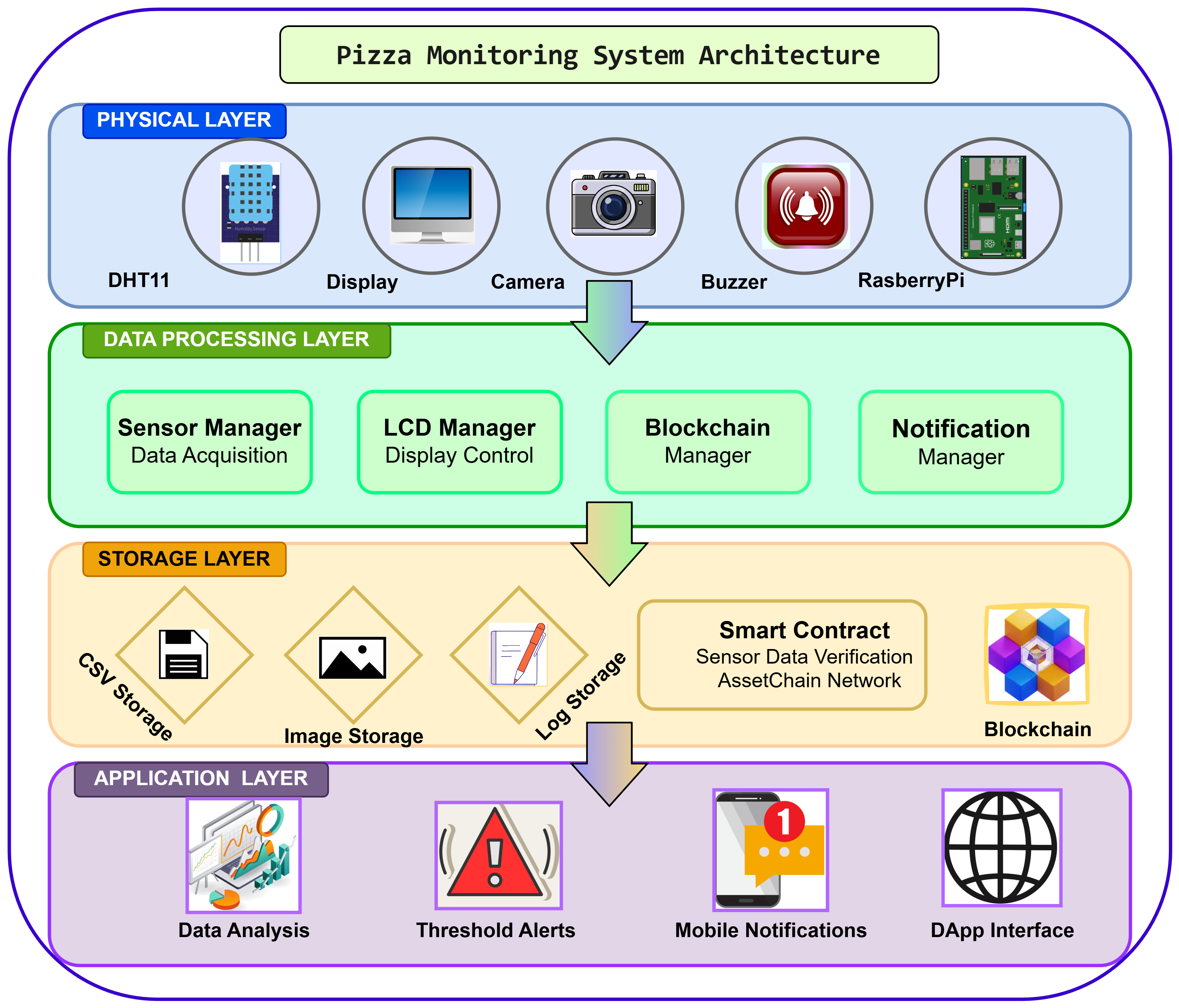}
  \caption{System Architecture of the IoT and Blockchain-Based Pizza Monitoring System.}
  \label{fig:H2sdiagram} 
\end{figure}

The architecture (Figure~\ref{fig:H2sdiagram}) is structured in four integrated layers. At the base, the \textbf{Physical Layer} includes key hardware components such as the DHT11 sensor for temperature and humidity monitoring, the Camera Module 3 for high-resolution image capture, a buzzer for local alerts, an LCD for on-site feedback, and a Raspberry Pi 4B that serves as the central processing and control unit. This layer handles data collection and initial processing. The \textbf{Data Processing Layer} interprets sensor raw input into actual data, identifying threshold breaches and enabling automated responses. It provides the intelligence behind real-time alerting and supports operational decision-making. Data is then secured in the \textbf{Storage Layer}, which leverages the Assetchain blockchain. Here, smart contracts handle the encrypted storage of sensor readings and metadata, ensuring records remain immutable and tamper-proof. Finally, the \textbf{Application Layer} delivers a user-friendly dashboard that visualizes live environmental data and historical trends. It also integrates a Telegram-based notification system to immediately alert staff of any critical conditions, even when they are not on-site. This multi-layered design delivers scalable and practical environmental monitoring for the food service sector.

The system collects sensor data every 30 seconds. Under normal conditions, the readings are recorded locally every 20 minutes. If a temperature or humidity threshold is breached for over 40 minutes, the buzzer activates, data is written to the blockchain, and a Telegram alert is sent. During critical events, image capture is triggered and logging frequency increases to 30 seconds. A daily CSV report and its cryptographic hash are uploaded to the blockchain to verify the authenticity of the data. This ensures a robust process for detecting anomalies and maintaining compliance with food safety standards.

Implemented in Solidity on Assetchain, smart contracts automate data logging, validate threshold breaches, and support batch data uploads. They include access controls and use compact data types for efficiency. Sensor data is securely indexed, ensuring traceability and low-cost blockchain interactions. The system’s blockchain layer ensures immutability and transparency, preventing tampering and enabling trusted audits. Data access is restricted through MetaMask authentication. Periodic blockchain updates and hashed reports reinforce data integrity without incurring high storage costs.
A static security analysis using Slither revealed no vulnerabilities, warning flags, or optimization issues. The contract code is lightweight (119 lines, 5 functions)and free from assembly use, making it robust and auditable. This system offers a resilient, secure, and scalable solution suited to pizza production. By combining the real-time insights of the IoT with the integrity of the blockchain data, it reduces food waste, improves safety, and builds trust in a fast-paced restaurant environment.

\section{Results and Analysis}
\label{result}
This section presents the evaluation of the IoT and blockchain-based monitoring system deployed in a working pizza shop. The analysis covers system accuracy, responsiveness, energy efficiency, and cost-effectiveness, validating its suitability for real-world food safety applications.

The system monitored refrigerator temperature and humidity within a defined range of $±2\,^\circ \text{C}$ to $6\,^\circ \text{C}$. The DHT11 sensor achieved an accuracy of $\pm$0.5°C for temperature and ±2\% for humidity. Blockchain integration enabled 100\% successful data uploads, with daily CSV reports and instant critical event logging. Recovery after system interruptions was also 100\% reliable. The system effectively detected threshold breaches and triggered real-time alerts via a buzzer, LCD message, and Telegram notification. Critical alerts were logged on the blockchain with a timestamp and followed by resolution updates. Notifications were consistently delivered within 2 to 4 seconds, ensuring a timely response from the staff. Power analysis showed the Raspberry Pi 4B as the highest consumer (2,500–6,000 mW). Optimizations such as limiting the DHT11’s active time and dimming the LCD reduced sensor and display energy use by up to 96\%. These measures make the system suitable for energy-constrained environments. Sensor data visualizations revealed stable daily conditions, with temperature/humidity spikes during peak hours due to frequent fridge access. Despite fluctuations, conditions remained within safe limits, preserving the quality of the ingredients. Using the AssetChain Testnet, the system averaged 12 transactions daily. The total daily cost was \$0.0107, with an estimated annual blockchain cost of \$3.91. This confirms the financial viability of secure, tamper-proof data storage in restaurant environments.

\section{Discussion and Conclusion}
\label{discussion_conclusion}

The proposed IoT and blockchain-based system significantly enhances food safety monitoring compared to traditional manual methods. By automating temperature and humidity logging and enabling real-time alerts, the system reduces human error and improves compliance with safety regulations. Daily reports are automatically generated and sent to the owner, creating a reliable and auditable record of environmental conditions. The integration of blockchain technology ensures tamper-proof, transparent data storage, reinforcing trust and accountability. Transaction analysis demonstrated that the system operates securely at a low cost, making it viable for small to medium-sized restaurants. In addition, energy-efficient strategies, such as duty-cycled sensors and LCD control, further improve its sustainability. Despite its effectiveness, some limitations remain. The system relies on the Raspberry Pi for processing and requires internet connectivity for real-time updates, which may not be ideal in all contexts. However, its affordability, scalability, and ease of deployment make it a practical solution for improving food safety and operational efficiency. Looking ahead, future improvements could focus on enhancing sensor precision, optimizing power consumption, and extending the system’s applicability to other sectors such as agriculture, logistics, and pharmaceuticals. This work presents a foundation for secure, transparent, and sustainable environmental monitoring, with the potential to impact a wide range of industries beyond pizza production.

\section{Acknowledgment}
This work was partially supported by the Italian MUR under D.M. 118/2023 for the PhD Program in Blockchain and DLT PNRR, the Project ECS0000038 “eINS - Innovation Ecosystem for Next Generation Sardinia”, CUP F53C22000430001, and by the AISAC Project, CUP B29J23001120005.

\bibliographystyle{plain}
\bibliography{biblio}

\end{document}